\documentclass[12pt,a4paper]{iopart}
\usepackage{iopams}
\usepackage{graphicx}

\newcommand{\eg}{e.\,g.}
\newcommand{\msun}{{\rm M}_{\odot}}
\begin{document}



\review[Multi-messenger approaches to binary SMBHs with PTAs]{Multi-messenger approaches to binary supermassive black holes in the ``continuous-wave'' regime}

\author{Sarah Burke-Spolaor}

\address{Jet Propulsion Laboratory, California Institute of Technology, M/S 138-310, 4800 Oak Grove Drive, Pasadena CA, 91109}
\ead{sarah.burke-spolaor@jpl.nasa.gov}
\begin{abstract}
Pulsar timing arrays are sensitive to gravitational waves from supermassive black hole (SMBH) binaries at orbital separations of $\ll1$\,pc. There is currently an observational paucity of such systems, although they are central figures in studies of galaxy evolution, merger dynamics, and active nucleus formation. 
We review the prospects of detecting SMBH binaries through electromagnetic radiative processes thought to be associated with galaxy mergers and late-stage binary evolution. 
We then discuss the scientific goals of joint pulsar timing and electromagnetic studies of these systems, including the facilitation of binary parameter estimation, identifying galactic hosts of gravitational wave emitters, and relevant studies of merger dynamics and cosmology. The use of upcoming high-precision timing arrays with the International Pulsar Timing Array and the Square Kilometre Array, combined with ongoing electromagnetic observing campaigns to identify active SMBH binaries, provide generous possibilities for multi-messenger astrophysics in the near future.
\end{abstract}
\submitto{\CQG}

\section{Introduction}
We are entering an exciting time for pulsar and gravitational wave (GW) astronomy, as upper limits on GWs from pulsar timing array (PTA) experiments are nearing the predicted level of emission from the population of binary supermassive black holes (SMBHs) in the Universe \cite{pptascience}. 
Binary SMBHs (mass $>10^6\,\msun$) have a principal role in leading theories of galaxy formation, in which massive galaxies were built up via the hierarchical merger of smaller galaxies within dark matter haloes \cite{VHM}. Single SMBHs reside at the center of most massive galaxies, and during a major merger SMBHs will form a binary in the merging system. During this process, the SMBHs will experience prolonged interaction with ambient stars and gas as the pair evolves.

Small-orbit binary SMBHs are estimated to be the strongest GW emitters detectable in the low frequency (nanohertz to microhertz) spectrum by pulsar timing experiments \cite{jenetPPTA}. As such, these objects represent the primary target of current PTAs \cite{nanograv,PPTA,epta}. For a general description of PTAs please refer to \cite{ipta}. 

Gravitational emission from SMBH binaries in the PTA frequency band is manifested in three flavours: ``continuous-wave'' emission (quasi-sinusoidal waveforms during steady inspiral), burst sources (e.\,g. as emitted at periastron of a high-eccentricity binary), and GW memory (a permanent change in space-time precipitated at binary coalescence; see J. Cordes article in this issue).
All of these GW emissions may contribute to a GW background.
However, sufficiently massive or nearby systems may rise above the stochastic background signal and be individually detectable.



In this article, we review emission processes that accompany SMBH binary systems in the continuous-wave regime detectable by pulsar timing, and assess the prospects of performing multi-messenger studies of these systems. Section \ref{sec:prelims} briefly reviews binary orbital parameters and GW detection in the pulsar timing frequency band. 
Section \ref{sec:fingerprints} reviews the electromagnetic emission that may accompany PTA targets.  
We discuss synergies between gravitational and electromagnetic wave studies of SMBH binaries in Section \ref{sec:symbiosis}, and inspect the prospects of finding and studying a multi-messenger target in Section \ref{sec:prospects}. Finally, we summarize the main conclusions of this paper in Section \ref{sec:conclusions}.

\section{Preliminaries}\label{sec:prelims}
An isolated (i.\,e.~not significantly influenced by external factors) SMBH binary, at the stage of PTA continuous-wave detection, can be described completely by 11 parameters:
sky position; 
primary and secondary SMBH mass $m_1, m_2$ (often conveniently combined as chirp mass, \mbox{$M_{\rm c}^{5/3}=m_1\,m_2\,(m_1+m_2)^{-1/3}$});
orbital period in the proper rest-frame of the binary, $P_{\rm bR}$;
time-varying orbital phase $\theta(t)$;
orbital inclination and line of nodes orientation, $i$, $\phi$, respectively;
eccentricity $e$ and the argument of periapsis; 
and co-moving distance $D$.
In an evolving system (i.\,e.~a binary significantly shrinking due to GW emission over the timescale of observation), $P_{\rm bR}$ and $e$ will be functions of time. Also note the relationship between the binary period observed at Earth, $P_{\rm b} = P_{\rm bR}(1+z)$, and the GW frequency detected in pulsar timing residuals, $f_{\rm g}=nf_{\rm b}$, where $n=2$ for circular orbits, or higher harmonics for eccentric orbits \cite{enoki}.

The orbit-averaged GW strain magnitude from a binary is given by \cite{jaffebacker}:
\begin{equation}\label{eq:hs}
h_{\rm s} = 4\sqrt{\frac{2}{5}}~\frac{(G\,M_{\rm c})^{5/3}}{c^4\,D}\Bigg(\frac{2\pi}{P_{\rm b}}\Bigg)^{2/3}~.
\end{equation}
Pulsar timing involves the precise tracking of pulse times-of-arrival by the modelling of known factors that influence their arrival times at Earth. For the best-timed pulsars, the RMS deviations in the arrival phase (``residuals'' after model removal) are on the order of a few tens of nanoseconds \cite{PPTA}. The time-of-arrival perturbation induced by a GW is given by the time-integrated full expression for GW strain, such that the residual scales with $h_{\rm s}/f_{\rm g}$, i.\,e. the residual $R \sim (M_{\rm c}^{5/3}/D)\cdot f_{\rm g}^{-1/3}$ \cite{wahlquist87}. The range of GW periods accessible to pulsar timing is defined by the total time span of the experiment and the observing cadence (currently $T\sim5$-20\,years, and $C\sim2$-4\,weeks, respectively, for the world-leading PTAs \cite{nanograv,PPTA,epta}).

Pulsar timing is sensitive to two GW-induced perturbations: that of the Earth-local space-time, which will exhibit a quadrupolar signature in pulsars distributed across the sky, and that of the pulsar-local space-time, which will typically be uncorrelated between pulsar pairs. 
The form of the latter perturbation, the so-called ``pulsar term'', is difficult to predict because pulsar distances are in general not accurately known and the geometric time delay between Earth and pulsar terms may allow source evolution to be evident in the pulsar term. 
However, given sufficient source evolution (i.\,e. the pulsar term signal is in a different frequency bin than the Earth term signal) the search for and use of the pulsar-term signal can raise PTA sensitivity, break $M_{\rm c}$--$D$ degeneracies to allow the measurement of cosmological-model-independent source distance Together with a sufficiently large signal-to-noise signal detection of GWs, it can allow evolution and higher-order orbital terms e.\,g. spin-orbit coupling to be estimated, motivating the development of algorithms sensitive to this signal \cite{paperfromarxiv,ellisMF,lee11,mingarelli12}.

There are a variety of algorithms that have been developed to search for continuous-wave signals in PTA data. These include Lomb-Scargle periodogram analyses that seek periodicities in the timing residuals \cite{jenet3c66b,yardleyetal10}, matched filter techniques \cite{ellisMF}, and a number of Bayesian or maximum likelihood methods which can identify single \cite{anholmCW,justinCWFstatistic} or multiple binaries \cite{babakCWmultiple}, and in some cases can estimate binary parameters (NB see article by J. Ellis in this issue, which details detection and parameter estimation).

\section{Electromagnetic SMBH Tracers from Host Merger to Black Hole Recoil$^\dag$}\label{sec:fingerprints}
\footnotetext{The processes sketched here are laid out in detail in various literature reviews (see, e.\,g.~\cite{begelman80,milomerritt05}).} Initial encounters between two SMBH-hosting galaxies cause disruption of galactic material, generating tidal tails and asymmetric morphologies that slowly diffuse as the central cores cease to be distinct pairs. Dynamical friction against stars and gas drives the SMBHs efficiently ($\lesssim$1\,Gyr) to the center of a merger remnant, where the black holes form a binary. During this time, gas will also flow into galactic central regions, and can trigger AGN activity and heightened star formation.
Stellar three-body interactions, gas accretion, and circumbinary disk formation can remove orbital energy from the binary, causing it to shrink. This phase of the inspiral can cause typically sharp central stellar cusps to flatten, inducing a visible ``mass deficit'' that is viewed as a shallow core profile \cite[e.\,g.]{merritt06,graham04}.
Energy dissipation in late-evolution binaries should be dominated by the emission of gravitational radiation. At coalescence, an asymmetric burst of GWs may induce a recoil that temporarily or permanently launches the coalesced black hole out of its central galactic position (with velocities up to hundreds or thousands of km\,s$^{-1}$; e.\,g.~\cite{campanelli07}).

\begin{figure*}
\begin{centering}
\includegraphics[width=1.0\textwidth,trim=17mm 28mm 16mm 19mm,clip]{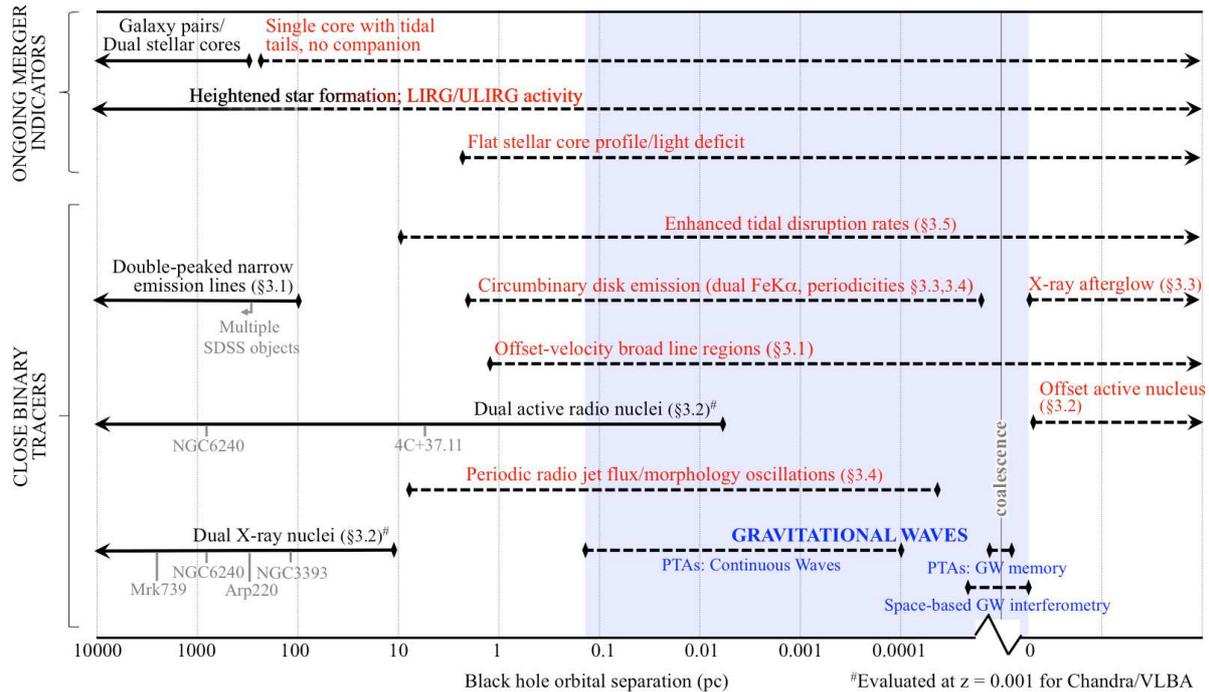}
\vspace{-6mm}
\linespread{0.8}
\caption{Here we display emission that may occur with SMBHs separated at scales shown on the x-axis. Emission from large-scale galactic processes (top), and emission that may arise from material directly influenced by a SMBH binary (bottom) are indicated separately. The red text/dashed lines indicate a signature that has not been detected, or that has been detected but not conclusively shown to mark a binary/recent merger. The shaded blue band indicates separations at which detectable GWs may be emitted. The lower spatial limit of resolved dual AGN (X-ray, radio) reflect the approximate resolution limits of current instruments (Chandra, VLBA) for $z=0.001$ targets, rather than a theoretical lower limit to the emission. Published, well-evidenced binary SMBHs separated below 3.5\,kpc are indicated with their discovery method (SDSS objects, \cite{liuconfirm,fuconfirm,shenconfirm}; Mrk\,739, \cite{koss11}; NGC6240, \cite{colbert1994,komossaetal03}; Arp220 \cite{arp220}; NGC3393, \cite{fabbiano11}; 4C+37.11, \cite{rodriguezetal06}).
\vspace{-4mm}\label{fig:bigone}
}
\end{centering}
\end{figure*}

Figure \ref{fig:bigone} demonstrates the complementarity of gravitational and electromagnetic waves as a funciton of the distance between the SMBH binary. 
Few $<$0.5\,kpc-separation systems have yet been discovered however searches are ongoing, as are efforts to corroborate observed indirect signatures of a binary's presence (\eg\ \cite{tingay11,liuinvestigation}).
The emissions detailed below all represent tracers of binary inspiral caused directly by SMBH energy outflows or tidal forces. These could alone be strong binary indicators, or provide a way to measure various binary parameters.
For each, we review when and how the signature might arise, and its observational status/prospects. For comparison with PTA efforts, we also estimate what parameters may be derived from the EM emission. 


\subsection{Peculiar AGN emission lines in UV/optical bands} 
The most recent attention has been on abnormal emission from the narrow and broad emission line regions fuelled by an active nucleus (e.\,g.~\cite{comerfordetal09,borosonlauer09,eracleous12}). If two black holes are active and in dominantly Keplerian orbits, their relative line-of-sight velocities can be estimated by
\begin{equation}\label{eq:dv}
	\delta v = \sqrt{\frac{G(m_1+m_2)}{a}}\sin i \cos \theta~,
\end{equation}
where $a$ is the orbital semi-major axis. In the binary scenario, broad emission lines emitted by SMBH-bound gas, or narrow emission lines emitted by unbound nuclear gas, may be red- or blue-shifted with respect to the host galaxy or the companion black hole by several hundred km\,s$^{-1}$.

At early stages of inspiral, two distinct narrow emission lines may remain intact, giving rise to double-peaked narrow lines. Such emission has been a successful indicator of a number of $a\gtrsim 0.5$\,kpc 
SMBH pairs \cite{comerfordetal09,liuconfirm,shenconfirm}, however such discoveries are limited to SMBH separations at which narrow line regions remain distinct (several hundred pc). These systems have been confirmed as genuine SMBH pairs through multi-wavelength follow-up efforts \cite{liuconfirm}. At PTA-relevant separations (well below 1\,pc), where the binary will exist within the extent of the narrow-line-emitting region, two effects may be visible: one may witness velocity-skewed or offset broad emission lines \cite{tsalmantza11,eracleous12}, or peculiar flux ratios between broad emission lines with different ionization potential from gravitational perturbations in the binary \cite{montuori12}.
Discoveries have been made of offset broad-line emission, however verification that the emission represents a genuine binary are forthcoming \cite{eracleous12,shen13}. Care must be taken with the interpretation of emission lines, as complex line emission dynamics may arise from a number of non-binary disk dynamical effects (\eg\ biconical outflows, \cite{shenconfirm}).
Yet another possibility is that offset broad emission lines can mark recoiling SMBHs \cite{eracleous12}.

The basic observable extracted from such emission is the line-of-sight velocity offset.
As indicated by Equation \ref{eq:dv}, it is expected that the binary velocity offset will grow as its orbit decays. We estimate that offset broad-line emission will be observable in binaries at separations of a few pc and below, down to an absolute lower limit of the Roche Lobe radius of the broad-line emission region (a few mpc \cite{eggleton83,mythesis}; however there is significant uncertainty in the size of the broad-line region). Simulations predict that asymmetric broad-line profiles will more commonly reveal smaller-orbit systems than double-peaked broad line profiles \cite{shenloeb}. If a binary is identified and periodic changes in velocity offset are tracked over time, the binary period can be directly inferred.


\subsection{Imaging double active nuclei}\label{sec:resolvedagn}
Two SMBHs in close orbit can be directly identified by spatially resolving characteristic SMBH tracers: either fluorescing Iron K$\alpha$ emission associated with an accretion disk (as seen in X-rays; \cite{komossaetal03}), or compact self-absorbed synchrotron emission at the site of relativistic jet formation (seen with a flat radio spectrum at 1--20\,GHz; \cite{rodriguezetal06}). In cases where only one emitter is discovered offset from the dynamical galactic centre, such emission may also indicate a recoiling SMBH rather than a binary \cite{volonteri08}.
Searches for spatially resolved binaries have mostly revealed wide-separation pairs \cite{komossaetal03,fabbiano11,3c75}, with the \emph{only} known sub-100\,pc pair, 0402+379, at a projected separation of 7\,pc \cite{rodriguezetal06}. In principle, such emissions could be sustained throughout evolution until accretion is quenched or the jet-formation mechanism is disrupted. In practice, the main current limitation on the smallest-resolvable binaries is the sensitivity and resolving power of current instruments. Chandra currently gives the best X-ray resolution at 0.5\,arcsec, and the excellent resolving power of radio interferometry enables down to $\sim$0.3\,milliarcsec resolution at 22\,GHz with the VLBA, corresponding to scales of 0.007\,pc at $z=0.001$ \cite{cosmocalc}. A standard $\Lambda$CDM cosmology has a peak in the angular-linear scale relationship at $z\sim1.6$, so that in principle the VLBA is sensitive to scales $>$2.8\,pc at \emph{all} redshifts at 22\,GHz. Large-scale blind searches are not practical for current X-ray or radio instruments given their small fields of view and the high sensitivity likely needed to unveil two SMBHs in a binary; however, such emission is an excellent tool for targeted confirmation of candidate binary systems \cite{radiocensus}. 

The direct observable using this emission is the projected pair separation $r_{\rm sep}$, i.\,e.~$a>r_{\rm sep}$. However, as in the case of peculiar emission lines, sustained observing of a resolved pair could directly measure its orbital period and semi-major axis, from which the system mass could be determined based on Keplerian motion.


\subsection{Circumbinary disk emission}
A circumbinary disk may form soon after a SMBH binary becomes a bound system. It is possible that in addition to the circumbinary disk, individual disks may exist around each black hole \cite[e.\,g.]{hayasaki08}. The binary can create a gap in the circumbinary disk, inside of which gas may accrete onto the SMBHs and outside of which accretion is subdued by tidal torques. When the GW emission becomes the dominant dissipator of orbital energy,
the disk undergoes viscous migration as it shrinks to follow the binary's inspiral. Eventually the binary coalesces, at which point the gas can evolve freely and reform an accretion flow. A number of emissions have been theorized to accompany this process:
1) Abnormally weak soft X-ray continuum and UV emission compared with a single AGN of similar mass \cite{tanaka12};
2) Double X-ray Fe\,K$\alpha$ emission lines from a pair of accretion disks \cite{sesanaXray};
3) Dips in the broad Fe\,K$\alpha$ emission lines from a circumbinary disk gap \cite{mckernan13};
4) After coalescence, renewed accretion may cause a gradually rising X-ray ``afterglow'' \cite{milosphinney,tanakamenou}. 
No targeted X-ray mission has yet been performed to identify any of these features as evidence of a binary, however a number of current (Chandra, NuSTAR, XMM) and future (IXO/Athena, Astro-H) X-ray satellites should be capable of detecting these with varying degrees of success (for instance, the MAXI detector may already be detecting periodic X-ray SMBH binary emission, while only a few double Fe\,K$\alpha$ lines are likely to be successfully probed by the upcoming Athena mission \cite{sesanaXray}; see individual papers for more detailed rate predictions).
In most cases, observing these signatures would generically indicate a recently or currently GW-emitting binary black hole at a given sky position, however would not provide specific parameters for PTA-based detection.

\subsection{Nuclear periodicities}
Here we describe two variants of periodic emission that might arise from binary SMBH systems: periodicities from orbital motion through a disk, and those caused by binary-induced jet precession.

Periodic emission is expected to arise from orbital passages through a gaseous disk that triggers temporarily heightened accretion or supersonic shocks in ambient gas \cite{bon12}. This is more likely in an asymmetric system, e.\,g. with a sufficiently high mass ratio or high-eccentricity orbit, where a secondary (less massive) SMBH may transit a circumbinary disk or an accretion disk around the primary black hole. Periodic emission at many wavelengths may result; X-ray or UV (with subdued optical/IR) periodicities may accompany ephemeral inflowing streams from the circumbinary disk \cite{hayasaki08,sesanaXray}, tidal torques on disks may cause optical variability (e.\,g.~in OJ\,287, \cite{sillanpaa88}), intensified radio core emission may accompany heightened accretion episodes (e.\,g.~PKS\,0637--752, \cite{godfrey12}), or shock waves from binaries within a single broad line region may produce variability in broad line emission (e.\,g.~NGC\,4151, \cite{bon12}). In principle, such emissions may commence after a SMBH pair forms a binary ($\sim$10\,pc), and continue until close to coalescence.

If a jet is being emitted from only one SMBH, it is possible that periodicities in the single jet may be caused by a binary. Precession of the jet may occur due to misaligned SMBH spin and orbital angular momentum vectors, or periodic variations in the direction and magnitude of the jet may be directly caused by the orbital motion of the black hole. In all cases, these variations will differ in period and opening angle from those due to simple geodetic precession \cite{kaastraroos92}. Sinusoidal and other periodic structures in many radio jets have been observed, however only a few have been attributed to a binary, with a notable few varying on the $P\lesssim20\,$yr timescales relevant to PTAs \cite{gower82,britzen10,nadia11}. One purported low-redshift binary, 3C66B, was reported to exhibit elliptical core motions with a period of $1.05\,$yr \cite{old3C66B}, however it was quickly noted that PTA data rules out the presence of the proposed binary because the strong GWs expected from such a system were not observed \cite{jenet3c66b}. 
\footnote{Newly-observed periodicities in 3C66B's $\lambda=3$\,mm flux have recently motivated a new, lower-mass, binary model \cite{new3C66B}.}

In nearly all cases of detectable periodicity we should expect that the observed periodicities should be 0.5 or 1.0 times that of the orbit, or potentially an integer harmonic thereof. For eccentric systems, the detection of two emission episodes per period (passages at either side of periastron) could allow an estimation of the orbital ellipticity. 


\subsection{Enhanced tidal disruption event rates}
Stars passing within the tidal disruption radius of a SMBH will be destroyed by tidal forces, which may result in X-ray, UV, optical, or radio outbursts \cite{donley02,gezari08,giannios11}. In a binary system, three-body slingshot interactions and loss-cone refilling effects can fuel tidal disruption event rates that are $10^2$--$10^4$ times higher than they would be for a single system of the same mass \cite{TDEchen,TDEivanov,TDEwegg}. This has led to the suggestion that in fact $\sim$3\% of all detected tidal disruption events should be occurring in late-stage binary SMBHs, and any object with multiple events detected is highly likely to be a binary \cite{TDEwegg}. This event excess should commence after dynamical friction ceases to be effective (i.\,e.~when three-body slingshot effects begin to extract orbital energy; $\lesssim$10\,pc), and endure until late stages of binary evolution. Tidal disruption excesses have also been predicted as a prompt response to SMBH recoil \cite{TDEstone}.

Tidal disruption events as an indicator of binary SMBHs may require further consideration after a more thorough assessment of the observable properties of these events. The events may also allow the detection of a large range of binary system masses (with primary masses down to $<$10$^6\,\msun$ \cite{TDEivanov}), thus many binaries identified in this way may not be resolvable continuous-wave sources for PTA experiments. So far, a few candidate tidal disruption events have been identified \cite{donley02,gezari08,vanvelzen11}, and current/upcoming high-sensitivity transient surveys (e.\,g.~LSST, PTF \cite{LSST,PTF}) should provide avenues to further explore such emission.



%
%


\section{The Symbiosis of Electromagnetic and PTA Studies of SMBH Binaries}\label{sec:symbiosis}
Electromagnetic and GW-focussed target studies of SMBH binaries or binary candidates can have significant interplay before and after the event of GW detection. Below, we note the relevant parameters of SMBH binary orbits, then discuss potential scientific goals in the pre- and post-GW detection eras.

\subsection{Before GW detection: Facilitating discovery and informing signal predictions}
PTA sensitivity to a SMBH binary would be enhanced by the electromagnetic measurement of any system parameters.
The efficacy of this improvement is algorithm- and PTA-dependent. Qualitatively, the improvement arises by effectively removing search parameters (e.\,g. the source position, which may be determined to high accuracy) or by limiting the parameter space that a binary may occupy (\eg\ through estimates of distance or system mass based on host galaxy properties), thus effectively lowering the false alarm rate. In practice, this enables a smaller prior volume in Bayesian methods, or fewer simulation/template trials for frequentist estimation or matched-template methods. 
At the very minimum, each of the electromagnetic markers detailed in Section \ref{sec:fingerprints} would provide a precise sky location. Sky position and GW frequency, which in addition to redshift/distance are the most accessible parameters to electromagnetic emission, introduce significant uncertainty in PTA false alarm estimation (e.\,g.~\cite{ellisMF}).


Electromagnetic studies may furthermore address significant unknowns in PTA searches, such as the influence of gas on the GW waveform during the inspiral phase, or how prevalent eccentricity is in PTA systems.
Some studies may allow actual orbits to be resolved (Section \ref{sec:resolvedagn}), revealing a full orbital template for PTAs to target with maximum-sensitivity matched-template detection methods \cite{ellisMF}.

PTA-derived \emph{upper limits} on GW parameters may also inform electromagnetic observations; in the absence of a detection, limits will be placed on a variety of binary parameters, or may completely rule out a binary model \cite[\eg]{jenet3c66b}. This could refine the binary interpretation, or warrant re-interpretation of the emission physics. This would be particularly interesting for the interpretation of excessive jet precession or heightened tidal disruption, where there are only limited or tenuous interpretations of such emission beyond a binary SMBH hypothesis.



\subsection{Upon GW detection: Host confirmation, parameter determination, and multi-messenger science}
When a quadrupolar signal is detected via pulsar timing, the conclusive identification of a host galaxy would provide decisive proof that the detected signal is indeed a GW from a SMBH binary. Symptomatic electromagnetic emission is crucial for host identification, particularly as the volumetric error for the first GW detection may be upwards of 2500\,deg$^2$ with a $\sim$80\% distance error (see the article by J.~Ellis in this issue), potentially containing millions of galaxies \cite{2df}. Given the large expected masses for PTA systems and a limited sky area, Figure \ref{fig:bigone} and Section \ref{sec:fingerprints} provide an excellent array of efficiently observable signatures (\eg\ galaxies with excess UV emission, core profile or galactic morphology surveys, optical spectroscopy), as well as opportunities to identify and track the SMBHs directly through high-sensitivity monitoring over orbital timescales (radio imaging, time-varying emission lines, circumbinary disk oscillations). In Section \ref{sec:emfraction} below, we estimate that 100\% of PTA-discovered systems should have large-scale electromagnetic identifiers, while $\sim$30\% may have actively-emitting nuclei.

Electromagnetic limits on any binary system parameters will assist in the extraction of the remaining parameters from PTA data. This may allow for some novel multi-messenger studies; for example,
photometric or spectroscopic measurement of the host's cosmological redshift would allow a distance estimation for a binary, enabling a chirp mass to be derived from an Earth-term-only PTA detection even for a SMBH binary without frequency evolution over PTA observing timescales. This would provide an interesting calibrator for SMBH mass-to-host relations \cite{ferrarese00,haring04} and address how mergers may form or skew these relations. 
The use of the pulsar term to detect source evolution and break degeneracies between $M_{\rm c}$ and $D$ (again, note predictions in the article by J. Ellis in this issue) would enable the use of the GW as a ``standard siren'' and a new cosmological distance measure if the host is known, as was originally suggested for the LISA satellite \cite{bloomwhitepaper}.
Looking again to an ideal future, if electromagnetic observation were to produce a precise signal template accessible by pulsar timing, it might finally enable a full pulsar and Earth-term fit that includes the detection of precise pulsar distances \cite{ellisMF}. This would provide the most accurate published stellar distances for all pulsars in the array (accurate to one GW wavelength, $\lambda\sim0.01$--6\,pc). This would render these pulsars precise calibration tools for galactic electron density models and planetary ephemerides \cite{ne2001,verbiestetal08}.

Any identification of a SMBH binary host could allow studies of the expedience of SMBH inspiral versus the galactic relaxation timescale, AGN triggers in merging galaxies, and SMBH growth through merger-induced accretion.
Multi-frequency monitoring of a binary host may also provide us with unexpected electromagnetic markers of GW systems which would incite exploration of unmodelled nuclear physical processes. This could feed into other PTA target identifications or inform future space-based GW interferometers of late-inspiral binary markers.

\begin{figure*}
\begin{centering}
\includegraphics[angle=270,width=0.75\textwidth,trim=0mm 0mm 0mm 0mm,clip]{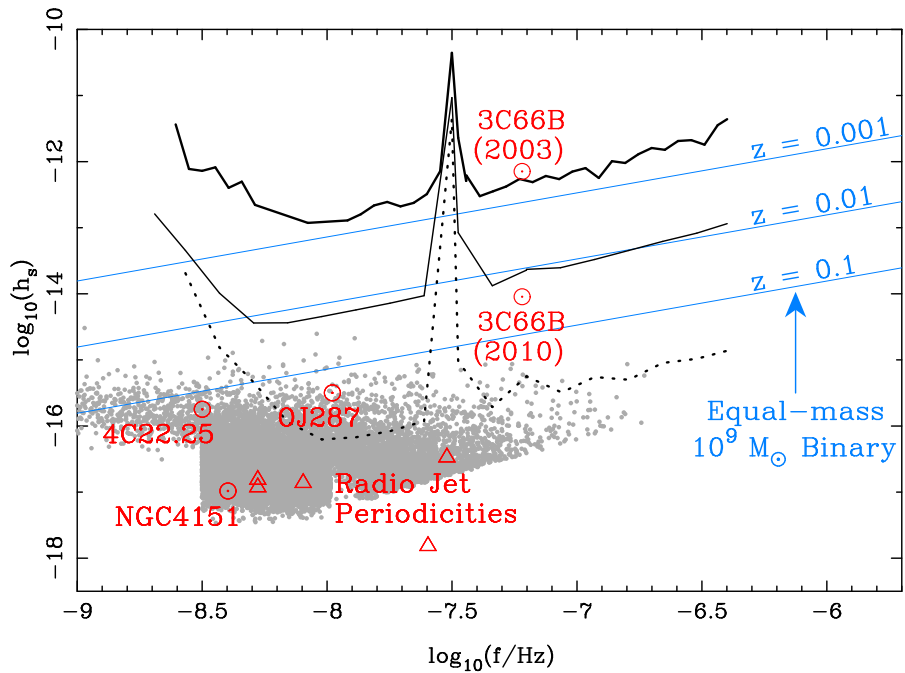}
\vspace{-2mm}
\caption{Here we show GW strain, $h$, as a function of GW frequency, $f_g$. The $h_{\rm s}$ of a $m_1=m_2=10^9\,{\rm M}_\odot$ SMBH binary at various redshifts is indicated. The thick solid curve indicates the most recently published PTA limit \cite{yardleyetal10}, and the thin solid curve indicates the estimated sensitivity of coherent techniques applied to the IPTA ($N_{\rm p}=40$, $T=15\,$yr). A future PTA timed with the SKA is shown by the dotted line ($N_{\rm p}=100,$ $T=10\,$yr, RMS residuals $=$ 20\,ns). The peak at $f_g=1\,{\rm yr}^{-1}$ is a lower limit, as GW signals at this frequency are completely absorbed by pulsar position fits.
Grey points indicate a standard simulated SMBH binary population from hierarchical cosmology (as in \cite{sesanaXray}; A. Sesana, private comm.).
We have estimated $h_{\rm s}$ for actual candidate SMBH binaries as marked. For all of these, we assumed representative parameters from the ranges presented in the literature. All were assumed to have circular orbits for placement (Refs: 3C66B \cite{old3C66B,new3C66B}; OJ287 \cite{sundeliusOJ287}; NGC4151 \cite{bon12}; jet periodicities as triangles, e.\,g.~\cite{britzen10,nadia11}; 4C22.25 \cite{decarli}). With the exception of OJ287 which has precisely estimated parameters, all candidates' placement within this phase is estimated from only partial information on SMBH masses, period, and orbital parameters. Without further observational refinement of these parameters, each object has an error box of up to several orders of magnitude in $h_{\rm s}$ and $f_{\rm g}$. 
}\vspace{-4mm}\label{fig:sensitivity}
\linespread{1.0}
\end{centering}
\end{figure*}

\section{Prospects for the Detection of EM+GW Continuous Wave Sources}\label{sec:prospects}
Here we assess the prospects for multi-messenger astrophysics with current and upcoming PTAs, considering the circumstances that will influence the feasibility of binary detectability in both the gravitational and electromagnetic wave domains.

\subsection{The number and population of PTA resolvable continuous-wave sources}
The SMBH background is expected to be constructed by millions of binary SMBHs throughout the local Universe; numerous predictions for the SMBH GW background detectable through PTA methods agree that the primary signal contributors are binaries with $z<2$ and $M_{\rm c}\gtrsim10^8\,\msun$  \cite[\eg]{jaffebacker,sesanagwb}. 

For a single SMBH binary to be ``resolvable'' above the background ensemble of signals, the source must exceed the confusion limit per frequency bin of size $C^{-1}-T^{-1}$, where the total number of bins is $F= T\cdot(C^{-1}-T^{-1})$ \cite{boylepen}. The per-bin confusion limit in an infinite signal-to-noise regime has been derived as $2N_{\rm p}/7$, in principle allowing an $N_{\rm p}$-pulsar array to resolve $2N_{\rm p}F/7$ binaries, i.\,e.~tens to hundreds, depending on the duration of and number of pulsars in the timing array \cite{boylepen}.\footnote{This estimate is valid only for pulsars with distances known to roughly a GW wavelength, which is not yet realistic for a full PTA (though may be in the future, \eg\ \cite{deller13}). For pulsars with poorly known distances, the authors note that there would be \mbox{$2N_{\rm p}F[1-(2F)^{-1}]/7$} resolvable binaries.} 

Of course, PTAs are not in the infinite signal-to-noise ratio regime. Thus, of more immediate importance is that $h_{\rm s}$ of an un-confused binary must exceed the strain detectable by a PTA. Based on estimations from standard hierarchical structure formation models (see the excellent Fig.~2 in \cite{sesanasingle}), the number of detectable binaries is currently $\leq$1 in a 5-year PTA using the $\sim$3 best-timed pulsars (50-200\,ns RMS residuals) in the International Pulsar Timing Array (IPTA\footnote{See \cite{ipta} and the article by D. Manchester in this issue regarding the status and capabilities of the IPTA.}). The expected numbers should, however, increase to 10's or more in the coming years as the IPTA becomes fully established, data sets grow longer, and ongoing surveys discover more pulsars that can be timed to high precision (\eg\ of the pulsars discovered in a current Arecibo survey, two were found to be suitable for PTAs \cite{deneva12,crawford12}).


\subsection{The fraction of electromagnetically-emitting PTA targets}\label{sec:emfraction}
If dynamical evolution theories are correct, \emph{all} major mergers should at some point be accompanied by large-scale merger indicators like central mass deficits or tidal tails, and heightened tidal disruption rates from incident stars. As long as these signatures endure long enough to accompany the epoch of PTA observation, they may at least allow the identification of a limited number of candidate binary host galaxies within an observed GW source position error box.

The other electromagnetic processes reviewed above require some level of gas inflow during merger to trigger the AGN and/or circumbinary disk activity that will give rise to the emission. The major question for potential multi-messenger targets is: how many PTA-detectable binaries will also have sufficient gas to allow their observation and/or tracking of binary system parameters?
Unfortunately, there is only sparse (and conflicting) information on the role of AGN emission during late stages of merger. Approximately 10\% of general galaxy samples have AGN activity, 
however some evidence has suggested that in a selection of \emph{early-stage mergers} (identified through spectroscopic galaxy pairs), the AGN fraction is up to $\sim$2.5 times higher than a control sample. Furthermore, the AGN fraction is systematically \emph{rising} for galaxies at closer separations down to the experiment's pair identification limit of $\sim$10\,kpc \cite{ellison11}. Here, we will take this as a crude argument that the fraction of mergers containing sufficient gas for electromagnetic nuclear activity may be $\gtrsim30$\%. This holds significant caveats, however; for instance, fuelling of one or both SMBHs is uncertain within a circumbinary disk \cite[\eg]{tanaka12,hayasaki08,arty96}, and the gas fraction in mergers may decrease at higher galaxy mass \cite{stewart09}.

\subsection{Current and future PTA detection prospects for electromagnetic binary candidates}
The most recently published sensitivity limit for continuous-wave emission in PTAs \cite{yardleyetal10}, which used an 18-pulsar, $T\sim10\,$year data set from Parkes Telescope, is shown as the thick solid line in Figure \ref{fig:sensitivity}. 
Several factors will already have significantly raised the sensitivity of pulsar timing, and so we will scale this curve in attempt to estimate the sensitivity of the IPTA and a future PTA with the Square Kilometre Array (SKA) to continuous-wave GW signals. As previously stated, GWs from SMBH binaries will show a quadrupolar correlation in pulsars located around the sky. Previous analysis methods did not use this information \cite{lommenbacker,jenet3c66b}, while recent ``coherent'' algorithms employ these correlations to maximize the sensitivity of PTAs. For instance, in comparing an incoherent spectral-power summing method as in \cite{yardleyetal10} to a matched filter technique, the $F_p$-statistic detection method reaches a factor of $\sim$10 times the sensitivity for the same data (J. Ellis, private comm.; see also \eg\ \cite{ellisMF}). 
The IPTA, when fully formed, will combine world-leading timing data for further sensitivity improvements, where the sensitivity should scale with number of pulsars as \mbox{$N_{\rm p}^{1/2}$}. Scaling from the previous 18-pulsar data set to the $\sim$40 IPTA pulsars \cite{ipta} and assuming similar RMS residual levels, this would allow a further factor of 1.5 boost in sensitivity. Additional factors of improvement will arise from longer data sets (the complete IPTA will include some pulsars with $\sim$20 year data spans), and smaller RMS residuals due to recent telescope backend improvements (e.\,g.~wider bandwidths and improved polarization calibration). These can likely add a further factor of two or more in sensitivity (\cite{xaviscaling}; X.~Siemens private comm.). As previously noted, electromagnetically-identified binary candidates with measured parameters will further heighten sensitivity to individual target systems.

As visible in Figure \ref{fig:sensitivity}, these compounded improvements greatly broaden the horizon within the reach of pulsar timing, and draw PTAs closer to the expected signal ranges of current binary candidates and predictions from hierarchical cosmology for binary SMBH populations \cite{sesanasingle,sesanaXray}. Treating electromagnetically-identified binary candidates as PTA targets,
those within closest reach appear to be OJ287 and 3C66B, based on revised parameters (3C66B was originally shown to lack its predicted GW signal; \cite{jenet3c66b}). However, given the significant uncertainty in the frequency, masses, and inclination of the other objects, the expected signal from these proposed binaries could be accessible by PTAs in the coming decade. The IPTA may already be able to place limits on the most extreme physical parameter models for some of these objects, and a PTA using the SKA may place upper limits on GWs that lead to refinement or reinterpretation of various systems (particularly OJ287, whose flux periodicity has led to a binary model out to a 3.5th-order post-Newtonian approximation \cite{crazyOJ287}). In fact, it appears that the SKA sensitivity may allow its timing array to reach the confusion limit of \cite{boylepen}, if the predictions of N-body hierarchical structure formation models are correct \cite{sesanaXray}. 
Furthermore, ongoing and upcoming searches electromagnetic binary signatures are exploring a significant discovery space even at redshifts $z<0.1$, and may yet reveal systems that are now or will soon be PTA-accessible.

\section{Conclusions}\label{sec:conclusions}
The main conclusions of this review are:
\begin{enumerate}
	\item \emph{Multi-messenger science is feasible with PTAs}. We estimate that 100\% of PTA-detectable orbiting SMBH binaries should have an identifiable host galaxy, and 30\% may have electromagnetically active SMBHs. 
	\item \emph{SMBH binaries have significant electromagnetic discovery potential}. Ongoing efforts aim to confirm some binary indicators and derive system parameters through radio imaging and time-dependent monitoring of abnormal emission lines or precessing jets. However, some theoretical markers of binary systems (circumbinary disk emission, tidal disruption excesses) have yet to be actively sought by observation. Targeted programs with current (VLBA, Chandra, NuSTAR) and upcoming (LSST, IXO/Athena, Astro-H, SKA) facilities have the potential to discover hundreds of binaries, many of which may lie within PTA spectral sensitivities.
	\item \emph{PTAs are already performing continuous-wave astrophysics}. Limits from pulsar timing already ruled out one proposed binary system \cite{jenet3c66b} and limited binaries in nearby galaxies \cite{lommenbacker}. The IPTA may contribute a sensitivity increase factor of more than 5 compared to the latest published PTA limit on continuous-wave emission from SMBH binaries, allowing it to access binaries of a given mass ratio out to a factor of 10 higher redshift. While the estimated IPTA sensitivity cannot yet reach the predicted GW emission from current binary candidates, it may be able to place meaningful astrophysical limits on several systems. Ongoing electromagnetic searches have sensitivity to massive systems within the PTA-accessible volume. As they discover new candidate systems, pulsar timing may be able to put strict limits on some of these, streamlining our understanding of emission near SMBH binaries.
	The predicted SKA sensitivity will detect or refute at least one candidate binary system already identified through its electromagnetic emission, and in fact may reach the confusion limit for continuous-wave targets.
	\item \emph{The future is looking bright}. The electromagnetic discovery of SMBH binaries has significant benefits before and after GW detection. Electromagnetically-measured parameters may raise PTA sensitivity enough to expedite the first GW discovery by constraining the parameter space probed by all PTA detection methods. After GW detection, electromagnetic parameter estimation will facilitate the GW-based measurement of full binary orbital parameters. The identification of a GW emitter's host through multi-wavelength surveys can lead to novel multi-messenger studies of black hole masses and cosmological distance measures. Concomitant observation of SMBH binaries will touch on a number of areas of astrophysics, including merger-induced accretion, AGN emission, and galaxy core/cusp formation.
\end{enumerate}

\ack
Part of this research was carried out at the Jet Propulsion Laboratory, California Institute of Technology, under a contract with the National Aeronautics and Space Administration. We thank our two referees for their useful comments on the manuscript.

\section*{References}
\bibliographystyle{unsrt}
\bibliography{cqg_article_burke-spolaor}

\end{document}